\documentclass[11pt,a4paper]{article}
\usepackage{jstyle}
\usepackage{amsthm,bm,mathrsfs,stmaryrd}
\usepackage[vcentermath]{youngtab}
\usepackage{simplewick}

\newcommand{\wh}{\widehat}
\newcommand{\wt}{\widetilde}

\author{Euihun JOUNG \quad}
\author{Karapet MKRTCHYAN}

\affiliation{Scuola Normale Superiore and INFN\\
Piazza dei Cavalieri 7, 56126 Pisa, Italy}

\emailAdd{euihun.joung@sns.it}
\emailAdd{karapet.mkrtchyan@sns.it}

\title{\centering Higher-derivative massive actions \\
from dimensional reduction}

\abstract{
A procedure to obtain higher-derivative free massive actions is proposed.
It consists in dimensional reduction of conventional two-derivative massless actions, where solutions to constraints bring in higher derivatives.
We apply this procedure to derive
the arbitrary dimensional generalizations of (linearized)
New Massive Gravity and New Topologically Massive Gravity.
}

\begin{document}

\maketitle

\section{Introduction}

Gravity as a theory of massive spin two particle
has been considered by many authors in view of
 its possible advantage in Cosmology.
However, a conceptual problem arises concerning the fate
of diffeomorphisms.
Indeed, the (linearized) diffeomorphism invariance is broken in the Fierz-Pauli (FP) massive action
(which describes a free massive spin two particle)
by the mass term.

\paragraph{New Massive Gravity}
In three dimensions,
there exists another description of free massive spin two
(with the linearized Einstein tensor equal to $-\frac12\, G_{\mu\nu}$):
\be\label{NMG}
	S_{\sst\rm NMG}[h_{\mu\nu}]=
	\int d^{3}x\left(
	R^{\mu\nu}\,R_{\mu\nu}-\frac38\,R^{2}
	-\frac{m^{2}}{4}\,h^{\mu\nu}\,G_{\mu\nu}\right),
\ee
whose nonlinear version, the so-called New Massive Gravity (NMG) \cite{Bergshoeff:2009hq,Bergshoeff:2009aq},
admits the diffeomorphism invariance.
Although this action is of fourth order in derivatives,
the theory is unitary due to the fact
that the (massless spin-two) ghost  mode does not propagate in three dimensions.
Moving from three to four dimensions,
it has been shown \cite{Bergshoeff:2012ud} that a similar mechanism
can be realized by a four-derivative action. The latter is formulated in terms of a field
$h_{\mu\nu,\rho}$ satisfying
\mt{h_{\mu\nu,\rho}=-h_{\nu\mu,\rho}}
and \mt{h_{\mu\nu,\rho}+h_{\nu\rho,\mu}+h_{\rho\mu,\nu}=0}\,.
A reasonable generalization of NMG to arbitrary $d$ dimensions
would involve a field of the hook symmetry:
\be\label{NMDG f}
	{\scriptsize \left.\yng(2,1,1,1)\right\}d-2}\,,
\ee
describing a massive spin two,
that is, the  {\tiny\yng(2)} representation of the massive little group $SO(d-1)$\,.

\paragraph{(New) Topologically Massive Gravity}

Besides the diffeomorphism invariant NMG,
the introduction of higher-derivative terms also allows
Lorentz-invariant description of a parity-violating single spin-two massive mode.
Topologically Massive Gravity (TMG),
the first action shown \cite{Deser:1981wh} to have such a spectrum,
is of third order in derivatives.
There is also a fourth-order model, so-called New Topologically Massive Gravity (NTMG)
\cite{Andringa:2009yc,Dalmazi:2009pm}, whose linearization reads
\be\label{NTMG}
	S_{\sst\rm NTMG}[h_{\mu\nu}]=\int d^{3}x\,
	\left(R^{\mu\nu}\,R_{\mu\nu}-\frac38\,R^{2}
	+\frac{m}4\,\epsilon^{\mu\nu\rho}\, h_{\mu}{}^{\sigma}\,
	\partial_{\nu}\,G_{\rho\sigma}\right).
\ee
The $d$-dimensional generalization of TMG has
also been found in \cite{Bergshoeff:2012yz}, and it concerns the (long) window representation:
\be\label{window}
	{\scriptsize \left.\yng(2,2,2,2)\right\} p}\in 2\,\mathbb N+1\,,
\ee 	
for the dimension \mt{d=2p+1}\,.
Hence, it exists only for \mt{{\rm mod}(d,4)=3}\,.

\medskip

Field theories with more than two derivatives are usually considered to be pathological,
as they generically contain ghost modes.
However, as one can see from the examples of NMG and (N)TMG
(see also \cite{Damour:1987vm,Deser:2009hb,Bergshoeff:2009tb,
Gullu:2010sd,Bergshoeff:2011pm,Ohta:2011rv,Hohm:2012vh}),
certain higher-derivative Lagrangians actually
describe unitary propagation, at least at the linearized level.
As we have briefly mentioned for NMG,
what underlies the unitarity of these theories
is that the ghost modes become pure gauge in certain dimensions.
This becomes possible by employing particular (unconventional)
off-shell fields for given dimensions, as the field \eqref{NMDG f}.
Several higher-derivative unitary theories have been constructed
relying on this property, while
the constructions have been, in our viewpoint, rather heuristic.
Having a more systematic way to derive higher derivative actions
might be useful in understanding and controlling
the (non-)unitarity of corresponding field theories.

In the present letter, we propose a systematic procedure to derive
a class of free higher-derivative massive theories with unitary propagation.
It is based on two observations: first, any conventional free massive theory can be
obtained from the corresponding massless one by dimensional reduction.
Second, the Hamiltonian constraints, inherited from the gauge symmetries
of the massless action, can be solved by substituting the fields with derivatives of
fields of other type (see e.g. \cite{Bekaert:1998yp,Henneaux:2004jw}).
Hence, massive actions with higher derivatives can be obtained
by the following procedure:
\begin{enumerate}
\item Begin with a $(d+1)$-dimensional massless action
in the Hamiltonian formulation,
where the conjugate fields are defined with respect to a spatial (say $z$) derivative
instead of the temporal one.
\item Solve the Hamiltonian constraints and substitute the (conjugate) fields with the corresponding solutions. This step increases the number of derivatives of the theory.
\item Perform dimensional reduction on the coordinate  $z$ to render the theory massive.
Use dualizations and on-shell equivalences to rewrite the action in different ways.
\end{enumerate}
The unitarity as well as the correct number of the degrees of freedom (DoF)
is guaranteed
since the propagating content is not affected at any step.
It is worth noticing that this procedure dictates  the particular
off-shell fields necessary for the non-propagation of ghost modes.
In principle, the procedure is general so applicable
to any kind of spectrum, but in this letter
we focus on two examples:
the $d$-dimensional generalizations of NMG and NTMG.\footnote{
In this paper, NMG and NTMG refer to the linearized versions of these theories.}

The derivation of NMG in any dimensions,
starting from the FP massless action, is presented in Section \ref{sec: NMG}.
In this case, we have a shortcut since a constraint appears already
in the Lagrangian formulation.
We perform dimensional reduction and solve the constraint.
This results in a fourth-order massive action with unconventional gauge symmetry.
Dualizing the field and using on-shell equivalence,
we get the $d$-dimensional generalization of NMG.
In Section \ref{sec: NTMG}, we turn to NTMG.
We first consider the topologically massive $p$-form field action,
whose construction shares the key features with that of NTMG.
In this construction, we eventually diagonalize the action
to recast it into two copies of topologically massive action.
As a byproduct, we also present the (anti-)self-dual massless action for these fields.
Finally, Section \ref{sec: disc} and Appendix \ref{sec: app}
contain respectively our conclusions and the derivation of some higher-derivative massless actions.

\section{New Massive Gravity}
\label{sec: NMG}

In this section, we show how the NMG action in $d$ dimensions  can be
obtained from the FP massless action in $d+1$ dimensions.

\subsection{Dimensional reduction}

The $(d+1)$-dimensional FP massless action reads (with $z:=x^{d}$)\footnote{
We use mostly plus signature.
The latin indices $m, n, \ldots$ run from $0$ to $d$\,,
while the greek ones $\mu, \nu, \ldots$ to $d-1$\,.
The (anti-)symmetrization of indices are with weight one. For example,
\mt{T_{(\mu\nu)}=\frac12\,(T_{\mu\nu}+T_{\nu\mu})\,,
	\ T_{[\mu\nu]}=\frac12\,(T_{\mu\nu}-T_{\nu\mu})\,.}}
\be\label{FP action}
	\mathcal S[\mathcal H_{mn}]=
	\frac{1}{4}\int d^{d}x\,dz\,
	\mathcal H^{mn}\,\mathcal G_{mn}\,,
\ee
where $\mathcal H_{mn}$ is the spin-two field (that is, the metric fluctuation)
and $-\frac12\,\mathcal G_{mn}$ is the corresponding (linearized) Einstein tensor.
The usual dimensional reduction
consists in fixing the $z$-dependence of the field as
\ba
	\mathcal H_{\mu\nu}(x,z)\e\sin(m\, z)\,h_{\mu\nu}(x)\,, \nn
	\mathcal H_{\mu d}(x,z)\e\cos(m\, z)\,h_{\mu d}(x)\,, \nn
	\mathcal H_{dd}(x,z)\e\sin(m\, z)\, h_{dd}(x)\,,
\ea
with the compactification \mt{z\in [0,2\pi/m]}.
After removing the $z$-dependence with the integration over $z$\,,
one ends up with the following massive action (modulo a factor of $\pi/m$)\,:
\ba\label{Sm}
	S[h_{\mu\nu},h_{\mu d}, h_{dd}]
	\e S_{\sst\rm E}[h_{\mu\nu}]
	-S_{\sst\rm M}[m\,h_{\mu\nu}
	-2\,\partial_{(\mu}h_{\nu)d}] \nn
	&&
	+\,\frac12\int d^{d}x\,
	h_{dd} \left(\partial^{\mu}\,\partial^{\nu}\,
	h_{\mu\nu}-\Box\,
	h^{\mu}_{\ \mu}\right)\,.
\ea
$S_{\sst\rm E}$ and $S_{\sst\rm M}$ are respectively the $d$-dimensional
FP massless action and the FP mass term:
\be
	S_{\sst\rm E}[h_{\mu\nu}]=\frac14\int d^{d}x\,
	h^{\mu\nu}\,G_{\mu\nu}\,,
	\qquad
	S_{\sst\rm M}[h_{\mu\nu}]
	=\frac14\int d^{d}x\left(h^{\mu\nu}\,h_{\mu\nu}
	-h^{\mu}_{\ \mu}\,h^{\nu}_{\ \nu}\right).
\ee
The action \eqref{Sm} admits the Stueckelberg symmetries:
\be\label{sym h}
	\delta\,h_{\mu\nu}=\partial_{\mu}\,\xi_{\nu}
	+\partial_{\nu}\,\xi_{\mu}\,,\qquad
	\delta\,h_{\mu d}=m\,\xi_{\mu}+\partial_{\mu}\,\xi_{d}\,,\qquad
	\delta\,h_{dd}=-2\,m\,\xi_{d}\,,
\ee
and, gauge fixing these symmetries, one gets
the FP massive action.

At this point, we take a non-standard way to proceed: we integrate-out the Lagrange multiplier
$h_{dd}$ instead of gauge-fixing it.
This gives the constraint:
\be\label{Const}
	\partial^{\mu}\,\partial^{\nu}\,h_{\mu\nu}
	-\Box\,h^{\mu}_{\ \mu}=0\,,
\ee
which can be solved as (see \cite{DuboisViolette:1999rd,DuboisViolette:2001jk,Bekaert:2002dt,Henneaux:2004jw})
\be\label{h(V)}	
	h_{\mu\nu}(\phi)=\partial^{\rho}\left(\phi_{\rho(\mu,\nu)}
	-\frac1{d-1}\,\eta_{\mu\nu}\,
	\phi_{\rho\lambda,}{}^{\!\lambda}\right),
\ee
where $\phi_{\mu\nu,\rho}$
satisfies \mt{\phi_{(\mu\nu),\rho}=0=\phi_{[\mu\nu,\rho]}}
having the symmetry of
\be
	\phi\ :\ {\scriptsize \yng(2,1)}\ .
\ee
After gauge-fixing the remaining $h_{\mu d}$\,,\footnote{
In fact,
the general solution of the constraint \eqref{Const}
also involves an arbitrary vector field $A_{\mu}$\,: \mt{h_{\mu\nu}\to
h_{\mu\nu}+\partial_{(\mu}A_{\nu)}}, but
it can be gauge-fixed together with $h_{\mu d}$\,. }
one ends up
with the following action for $\phi_{\mu\nu,\rho}$\,:
\be\label{FP V}
	S[\phi_{\mu\nu,\rho}]
	=S_{\sst \rm E}[h_{\mu\nu}(\phi)]
	-m^{2}\,S_{\sst\rm M}[h_{\mu\nu}(\phi)]\,.
\ee
This is nothing but the FP massive
action with \mt{h_{\mu\nu}=h_{\mu\nu}(\phi)} \eqref{h(V)}.
Although the novelty of this action vis-\`a-vis the FP one
seems to be trivial,
it actually encodes key information about NMG.
Indeed, we will show in the following that
the action \eqref{FP V} admits the gauge symmetries which are
equivalent to those of NMG.
This point will become manifest after dualization---the subject of the next section,
so let us conclude the present section by listing all
the gauge symmetries of the action \eqref{FP V}:
\begin{itemize}
\item
Firstly, the function $h_{\mu\nu}(\phi)$ itself is invariant with respect to
the following gauge transformation:
\be\label{theta sym}
	\delta\,\phi_{\mu\nu,\rho}
	=\partial^{\lambda}
	\left(\theta_{\lambda\mu\nu,\rho}
	-\frac13\,\theta_{\mu\nu\rho,\lambda}\right).
\ee
in the sense that \mt{h_{\mu\nu}(\phi)=
h_{\mu\nu}(\phi+\delta \phi)}\,.
The gauge parameter $\theta_{\mu\nu\rho,\lambda}$ is a tensor
totally antisymmetric in the first three indices
and satisfies \mt{\theta_{[\mu\nu\rho,\lambda]}=0}\,:
it has the symmetry of
\be
	\theta\ : \ {\scriptsize \yng(2,1,1)}\ .
\ee
\item Secondly, the action  \eqref{FP V} admits the gauge symmetry:
\be
	\delta\,\phi_{\mu\nu,\rho}=
	(\eta_{\rho\mu}\,\partial_{\nu}-\eta_{\rho\nu}\,\partial_{\mu})\,
	\sigma\,,
\ee
induced from the transformation \eqref{sym h}: it gives
\mt{\delta\,h_{\mu\nu}(\phi)
=\partial_{\mu}\partial_{\nu}\sigma}\,.
However, as we will see, this symmetry does not have
any counterpart in NMG.

\end{itemize}

\subsection{Dualization}

In order to see that the action $S[\phi_{\mu\nu,\rho}]$ coincides
with the NMG ones, we need to dualize the field $\phi_{\mu\nu,\rho}$\,.
The $GL(d)$ irreducible tensor
$\phi_{\mu_{1}\mu_{2},\nu}$ is dual to the direct sum of
two tensors  $\wh\varphi_{\mu_{1}\cdots\mu_{d-2},\nu}$
and $\wt\varphi_{\mu_{1}\cdots\mu_{d-1}}$\,:
\vspace{-3pt}
\be\label{phi dualize}
	{\scriptsize
	\overset{\ \displaystyle \phi\phantom{\Big|}}{\yng(2,1)}\ \simeq\
	\overset{\ \displaystyle \widehat\varphi \phantom{\Big|}}
	{\widehat{\yng(2,1,1,1)}}\Bigg\}{\sst d-2}
	\ \oplus\
	\overset{\hspace{-5pt} \displaystyle \widetilde\varphi \phantom{\Big|}}
	{\left.\yng(1,1,1,1,1)\right\}}{\sst d-1}}\,,
\ee
where the symbol $\widehat{\ \ }$ of Young diagram denotes
the traceless-ness.
Analogously, the gauge parameter
$\theta_{\mu_{1}\mu_{2}\mu_{3},\nu}$ is dual to
$\wh\vartheta_{\mu_{1}\cdots \mu_{d-3},\nu}$ and
$\wt\vartheta_{\mu_{1}\cdots \mu_{d-2}}$\,:
\vspace{-3pt}
\be\label{theta dualize}
	{\scriptsize
	\overset{\ \displaystyle \theta\phantom{\Big|}}{\yng(2,1,1)}\ \simeq\
	\overset{\ \displaystyle \wh\vartheta \phantom{\Big|}}
	{\widehat{\yng(2,1,1,1)}}\Bigg\}{\sst d-3}
	\ \oplus\
	\overset{\hspace{-5pt} \displaystyle \wt\vartheta \phantom{\Big|}}
	{\left.\yng(1,1,1,1,1)\right\}}{\sst d-2}}\, .
\ee
In the following, we specify these relations and show
that the dual action becomes that of NMG
after integrating-out $\wt\varphi_{\mu_{1}\cdots \mu_{d-1}}$\,.
We first consider the three dimensional case
for simplicity,
and then turn to the arbitrary dimensional case.

\subsubsection{Three dimensions}

Let us define the dual fields for the hook tensors $\phi_{\mu_{1}\mu_{2},\nu}$
and $\theta_{\mu_{1}\mu_{2}\mu_{3},\nu}$ as
\ba\label{Dual phi 3}
	&& \wh\varphi_{\mu\nu}+\wt\varphi_{\mu\nu}
	= -\frac12\,\epsilon_{\mu\rho\sigma}\,\phi^{\rho\sigma,}{}_{\nu}
	\qquad [\,\wh\varphi_{[\mu\nu]}=0=\wt\varphi_{(\mu\nu)}\,]\,,\nn
	&&\quad \vartheta_{\mu} =-\frac1{3!}\,
	\epsilon_{\rho\sigma\lambda}\,\theta^{\rho\sigma\lambda,}{}_{\mu}\,.
\ea
Note that in three dimensions, there is no
distinction between $\wh\vartheta$ and $\wt\vartheta$\,.
The inverse relations of the above formulas are
\ba\label{dual phi 3}
	&&\phi_{\rho\sigma,\nu}=
	\frac23\left(\epsilon_{\mu\rho\sigma}\,
	\wh\varphi^{\,\mu}{}_{\nu}-\epsilon_{\mu\nu[\rho}\,
	\wh\varphi^{\,\mu}{}_{\sigma]}\right)
	-\epsilon_{\mu\kappa[\rho}\,\eta_{\sigma]\nu}\,
	\wt\varphi^{\,\mu\kappa}\,,\nn
	&& \theta_{\rho\sigma\lambda,\nu}
	=3\,\epsilon_{\kappa[\rho\sigma}\,\eta_{\lambda]\nu}\,
	\vartheta^{\kappa}\,.
\ea
In terms of the dual fields, the gauge symmetry \eqref{theta sym} reads
\be\label{g tr 3}
	\delta\,\wh\varphi_{\mu\nu}=\partial_{(\mu}\,\vartheta_{\nu)}
	-\frac13\,\eta_{\mu\nu}\,\partial^{\rho}\,\vartheta_{\rho}\,,
	\qquad
	\delta\,\wt\varphi_{\mu\nu}=\partial_{[\mu}\,\vartheta_{\nu]}\,.
\ee

Let us remind the reader that the action $S[\phi_{\mu\nu,\rho}]$ \eqref{FP V} is given through
$h_{\mu\nu}(\phi)$\,\eqref{h(V)} and has two terms:
the four-derivative one $S_{\sst\rm E}[h_{\mu\nu}(\phi)]$
and the two-derivative one $S_{\sst\rm M}[h_{\mu\nu}(\phi)]$\,.
In the following,
we recast these two terms into functionals of the dual fields,
using the expression of $h_{\mu\nu}$\,:
\be\label{H phi 3}
	h_{\mu\nu}(\wh\varphi,\wt\varphi)
	=\epsilon_{\rho\sigma(\mu}\,\partial^{\rho}\,\wh\varphi^{\,\sigma}{}_{\nu)}
        +\frac12\,\epsilon_{\kappa\lambda(\mu}\,\partial_{\nu)}\, \wt\varphi^{\,\kappa\lambda}\,.
\ee

\paragraph{Four-derivative part}
Let us first consider the four-derivative part,
$S_{\sst\rm E}[h_{\mu\nu}(\wh\varphi,\wt\varphi)]$\,.
One can notice that
the $\wt\varphi_{\mu\nu}$ part of \eqref{H phi 3}
does not contribute since
it has a form of a gauge transformation \eqref{sym h} with the parameter \mt{\xi_{\mu}=\frac12\,
\epsilon_{\mu\kappa\lambda}\,\wt\varphi^{\kappa\lambda}}.
Hence, one ends up with a four-derivative action of $\wh\varphi_{\mu\nu}$
with gauge symmetry \eqref{g tr 3},
which is proportional to the \emph{Bach} action\footnote{
There is a unique four-derivative action for any field of two column Young diagram, which is invariant under the corresponding gauge and Weyl transformations. In this paper,
the latter will be referred as Bach action.}:
\be
	S_{\sst\rm E}[h_{\mu\nu}(\wh\varphi,\wt\varphi)]
	=S_{\sst\rm B}[\wh\varphi_{\mu\nu}]
	=S_{\sst\rm B}[\varphi_{\mu\nu}]
	:=\frac14\int d^{d}x\,G^{\mu\nu}\,S_{\mu\nu}\,.
\ee
Here $\varphi_{\mu\nu}$ is a traceful tensor whose
 traceless part is given by $\wh\varphi_{\mu\nu}$\,,
 and \mt{S_{\mu\nu}=G_{\mu\nu}-\frac12\,\eta_{\mu\nu}\,G^{\rho}{}_{\rho}} is the Schouten tensor.
 The second equality holds thanks to the Weyl symmetry of the Bach action.

\paragraph{Two-derivative part}
The two-derivative part, $S_{\sst\rm M}[h_{\mu\nu}(\wh\varphi,\wt\varphi)]$\,,
is given by
\ba\label{FP vb}
	&&S_{\sst\rm M}[h_{\mu\nu}(\wh\varphi,\wt\varphi)]
	= \frac1{4}\int d^{3}x\,\bigg[\,
	\big(\epsilon_{\rho\sigma(\mu}\,\partial^{\rho}\,\wh\varphi^{\,\sigma}{}_{\nu)}
	\big)^{2}
	-\big(\epsilon_{\rho\sigma(\mu}\,\partial^{\rho}\,\wh\varphi^{\,\sigma}{}_{\nu)}\big)
	 \big(\epsilon^{\kappa\lambda(\mu}\,\partial^{\nu)}\,
	 \wt\varphi_{\kappa\lambda}\big)
	\nn
	&&\hspace{140pt}
	+\,\frac14\,
        \big(\epsilon^{\kappa\lambda(\mu}\,\partial^{\nu)}\, \wt\varphi_{\kappa\lambda}
        \big)^{2}
        -\frac14\,\big(\epsilon^{\rho\sigma\lambda}\,\partial_{\rho}\,
        \wt\varphi_{\sigma\lambda}\big)^{2}\,\bigg]\nn
        &&=\,\frac1{4}\int d^{3}x\,\bigg[\,
	\big(\epsilon_{\rho\sigma(\mu}\,\partial^{\rho}\,\wh\varphi^{\,\sigma}{}_{\nu)}\big)^{2}
	-\big(\partial^{\rho}\,\wh\varphi_{\mu\rho}\big)
	\big(\partial_{\sigma}\,\wt\varphi^{\mu\sigma}\big)
	-\frac12\,\big(\partial_{\sigma}\,\wt\varphi^{\mu\sigma}\big)^{2}\,
	\bigg]\,.
\ea
The antisymmetric field  $\wt\varphi_{\mu\nu}$
can be integrated out from the above action by solving the $\wt\varphi_{\mu\nu}$-shell condition:
\be
	\partial_{[\mu}\,\partial^{\rho}\,\wt\varphi_{\nu]\rho}+
	\partial_{[\mu}\,\partial^{\rho}\,\wh\varphi_{\nu]\rho}=0\,,
\ee
as
\be
	\partial^{\rho}\,\wt\varphi_{\mu\rho}+
	\partial^{\rho}\,\wh\varphi_{\mu\rho}=
	\partial_{\mu}\,\chi\,.
	\label{wt os sol}
\ee
Here $\chi$ is an arbitrary field subjected to the condition:
\be
	\Box\,\chi =\partial^{\mu}\partial^{\nu}\wh\varphi_{\mu\nu}\,,
	\label{chi sol}
\ee
which is inherited from the antisymmetric property of $\wt\varphi_{\mu\nu}$\,:
\mt{\partial^{\mu}\partial^{\nu}\wt\varphi_{\mu\nu}=0}\,.
Hence, on the $\wt\varphi_{\mu\nu}$-shell,
the action \eqref{FP vb} becomes
\be
	\frac1{4}\int d^{3}x\,\bigg[\,
	\big(\epsilon_{\rho\sigma(\mu}\,\partial^{\rho}\,\wh\varphi^{\,\sigma}{}_{\nu)}\big)^{2}
	-\frac12\,\partial_{\sigma}\,\wh\varphi^{\mu\sigma}\,
	\big(\partial_{\mu}\,\chi-\partial^{\rho}\,\wh\varphi_{\mu\rho}\big)\,\bigg]\,,
\ee
with $\chi$ satisfying \eqref{chi sol}.
The latter condition on $\chi$ can be viewed as
an $\chi$-shell one resulting from an off-shell action $S_{2}[\wh\varphi_{\mu\nu},\chi]$\,.
Indeed, one can determine such an action as
\be\label{FP va}
	S_{2}[\wh\varphi_{\mu\nu},\chi]
	=\frac1{4}\int d^{3}x\,\bigg[\,
	\big(\epsilon_{\rho\sigma(\mu}\,\partial^{\rho}\,\wh\varphi^{\,\sigma}{}_{\nu)}\big)^{2}
	+\frac12\,\big(\partial_{\mu}\,\chi-
	\partial^{\rho}\,\wh\varphi_{\mu\rho}
	\big)^{2}\,
	\bigg]\,.
\ee
Let us notice that, as a consequence of \eqref{g tr 3} and \eqref{chi sol},
the field $\chi$ is subject to gauge transformation:
\be
 	\delta\,\chi=\frac23\,\partial_{\mu}\,\vartheta^{\mu}\,,
\ee
so one can consider the following
combination of $\wh\varphi_{\mu\nu}$ and $\chi$\,:
\be
	\varphi_{\mu\nu}:=\wh\varphi_{\mu\nu}+\frac12\,\eta_{\mu\nu}\,\chi\,,
\ee
that has a standard gauge transformation:
\be\label{st g h}
	\delta\,\varphi_{\mu\nu}=\partial_{(\mu}\,\vartheta_{\nu)}\,.
\ee
Therefore, the action \eqref{FP va} is a two-derivative functional
of $\varphi_{\mu\nu}$ with the gauge symmetry \eqref{st g h}.
This implies that it must be proportional to
the FP massless action. Indeed, we get
\be
	S_{\sst\rm M}[h_{\mu\nu}(\wh\varphi,\wt\varphi)]
	\approx S_{\sst\rm E}[\varphi_{\mu\nu}]\,,
\ee
where, by $\approx$\,, we mean the on-shell equivalence.

\medskip

Collecting the four- and two-derivative terms, one ends up with the action:
\be
	S_{\sst \rm E}[h_{\mu\nu}(\wh\varphi,\wt\varphi)]
	-m^{2}\,S_{\sst\rm M}[h_{\mu\nu}(\wh\varphi,\wt\varphi)]
	\ \approx\
	S_{\sst\rm B}[\varphi_{\mu\nu}]-m^{2}\,S_{\sst\rm E}[\varphi_{\mu\nu}]\,,
\ee
which is nothing but the linearization \eqref{NMG} of NMG \cite{Bergshoeff:2009hq}.

\subsubsection{General $d$ dimensions}

Arbitrary dimensional case is a straightforward generalization of the
three dimensional one, so we provide the formulas
parallel to the three dimensional ones, minimizing repetition of comments.

The field $\phi_{\rho\sigma,\nu}$ and the gauge parameter
$\theta_{\rho\sigma\lambda,\nu}$ are dualized,
instead of \eqref{Dual phi 3}\,, as
\ba\label{Dual phi d}
	&& \wh\varphi_{\mu_{1}\cdots \mu_{d-2},\nu}+
	\wt\varphi_{\mu_{1}\cdots\mu_{d-2}\nu}
	= -\frac12\,
	\epsilon_{\mu_{1}\cdots\mu_{d-2}\rho\sigma}\,
	\phi^{\rho\sigma,}{}_{\nu}\,,\nn
	&&\wh\vartheta_{\mu_{1}\cdots \mu_{d-3},\nu}
	+\wt\vartheta_{\mu_{1}\cdots\mu_{d-3}\nu}
	=-\frac1{3!}\,\epsilon_{\mu_{1}\cdots\mu_{d-3}\rho\sigma\lambda}\,
	\theta^{\rho\sigma\lambda,}{}_{\nu}\,,
\ea
where $\wt\varphi$ and $\wt\vartheta$ are totally antisymmetric tensors, while $\wh\varphi$ and $\wh\vartheta$ are hook-type ones with \mt{\wh\varphi_{[\mu_{1}\cdots \mu_{d-2},\nu]}=0=
\wh\vartheta_{[\mu_{1}\cdots \mu_{d-3},\nu]}}\,.
The inverse relations are given by
\ba\label{dual phi d}
	&& \phi_{\rho\sigma,\nu}=
	\frac{2}{3(d-2)!}\,\Big(
	\epsilon_{\mu_{1}\cdots\mu_{d-2}\rho\sigma}\,
	\wh\varphi^{\,\mu_{1}\cdots\mu_{d-2},}{}_{\nu}
	-\epsilon_{\mu_{1}\cdots\mu_{d-2}\nu[\rho}\,
	\wh\varphi^{\,\mu_{1}\cdots\mu_{d-2},}{}_{\sigma]}\Big)
	\nn
	&&\hspace{50pt}
	-\,\frac{2}{(d-1)!}\,\epsilon_{\mu_{1}\cdots\mu_{d-2}\kappa[\rho}\,
	\eta_{\sigma]\nu}\,
	\wt\varphi^{\,\mu_{1}\cdots\mu_{d-2}\kappa}\,,\nn
	&&\theta_{\rho\sigma\lambda,\nu}=
	\frac{3}{2(d-3)!}\left(\epsilon_{\mu_{1}\cdots\mu_{d-3}\rho\sigma\lambda}\,
	\wh\vartheta^{\,\mu_{1}\cdots\mu_{d-3},}{}_{\nu}
	-\epsilon_{\mu_{1}\cdots\mu_{d-3}\nu[\rho\sigma}\,
	\wh\vartheta^{\,\mu_{1}\cdots\mu_{d-3},}{}_{\lambda]}\right)\nn
	&&\hspace{50pt}
	+\,\frac{3}{(d-2)!}\,\epsilon_{\mu_{1}\cdots\mu_{d-3}\kappa
	[\rho\sigma}	\eta_{\lambda]\nu}\,
	\,\wt\vartheta^{\,\mu_{1}\cdots\mu_{d-3}\kappa}\,,
\ea
and the gauge symmetry \eqref{theta sym} becomes
\ba\label{dual phi sym d}
	\delta\,\wh\varphi_{\mu_{1}\cdots\mu_{d-2},\nu}
	\e(d-2)\,\bigg[\,
	\partial_{[\mu_{d-2}}\,\wh\vartheta_{\mu_{1}\cdots\mu_{d-3}],\nu}
	-\frac13\,\eta_{\nu[\mu_{d-2}}\,\partial^{\rho}\,
	\wh\vartheta_{\mu_{1}\cdots\mu_{d-3}],\rho}\nn
	&&
	+\,\partial_{[\mu_{d-2}}\,\wt\vartheta_{\mu_{1}\cdots\mu_{d-3}]\nu}
	-\partial_{[\mu_{d-2}}\,\wt\vartheta_{\mu_{1}\cdots\mu_{d-3}\nu]}
	-\frac13\,\eta_{\nu[\mu_{d-2}}\,\partial^{\rho}\,
	\wt\vartheta_{\mu_{1}\cdots\mu_{d-3}]\rho}\,\bigg]\,,\nn
	\delta\,\wt\varphi_{\mu_{1}\cdots\mu_{d-1}}
	\e-(d-2)\,\partial_{[\mu_{d-1}}\,\wt\vartheta_{\mu_{1}\cdots\mu_{d-2}]}\,.
\ea
Plugging \eqref{dual phi d} into \eqref{h(V)}, one gets
\be\label{H phi d}
	h_{\mu\nu}(\wh\varphi,\wt\varphi)
	=\frac1{(d-2)!}\,\epsilon_{\rho_{1}\cdots\rho_{d-2}\lambda(\mu}\,
	\partial^{\lambda}\,\wh\varphi^{\,\rho_{1}\cdots \rho_{d-2},}{}_{\nu)}
        +\frac1{(d-1)!}\,\epsilon_{\sigma_{1}\cdots \sigma_{d-1}(\mu}\,\partial_{\nu)}\,
        \wt\varphi^{\,\sigma_{1}\cdots\sigma_{d-1}}\,.
\ee
We now plug the above solution in the action \eqref{FP V}.
The antisymmetric field $\wt\varphi_{\mu_{1}\cdots \mu_{d-1}}$
does not contribute to
the four-derivative part $S_{\sst\rm E}[h_{\mu\nu}(\wh\varphi,\wt\varphi)]$
due to gauge invariance of the latter.
The resulting action is the Bach
action for the hook field $\varphi_{\mu_{1}\cdots\mu_{d-2},\nu}$\,:
\be\label{Bach2}
	S_{\sst\rm E}[h_{\mu\nu}(\wh\varphi,\wt\varphi)]
	= S_{\sst\rm B}[\varphi_{\mu_{1}\cdots\mu_{d-2},\nu}]
	:=\frac1{4\,(d-2)!}\int d^{d}x\,G^{\mu_{1}\cdots\mu_{d-2},\nu}\,S_{\mu_{1}\cdots\mu_{d-2},\nu}\,,
\ee
where $G^{\mu_{1}\cdots\mu_{d-2},}{}_{\nu}$ and
$S^{\mu_{1}\cdots\mu_{d-2},}{}_{\nu}$
are respectively the generalized Einstein and Schouten tensors given by
\ba
	&&G^{\mu_{1}\cdots\mu_{d-2},}{}_{\nu}=
	\epsilon^{\mu_{1}\cdots\mu_{d-2}\kappa\sigma}\,
	\epsilon_{\rho_{1}\cdots\rho_{d-2}\lambda\mu}\,
	\partial_{\kappa}\,\partial^{\lambda}\,\varphi^{\rho_{1}\cdots\rho_{d-2},}{}_{\sigma}
	\nn
	&&S^{\mu_{1}\cdots\mu_{d-2},}{}_{\nu}=
	G^{\mu_{1}\cdots\mu_{d-2},}{}_{\nu}
	-\frac12\,\delta^{[\mu_{1}}_{\nu}\,G^{\rho\mu_{2}\cdots \mu_{d-2}],}{}_{\rho}\,.
\ea
Let us notice that the Bach action \eqref{Bach2} has the Weyl symmetry
\mt{\delta\varphi_{\mu_{1}\cdots\mu_{d-2},\nu}=\eta_{\nu[\mu_{1}}\,\alpha_{\mu_{2}\cdots\mu_{d-2}]}}\,, where the parameter $\alpha_{\mu_{1}\cdots \mu_{d-3}}$ is a totally antisymmetric tensor.

On the other hand, the two-derivative part is given by
\ba\label{FPvb d}
	S_{\sst\rm M}[h_{\mu\nu}(\wh\varphi,\wt\varphi)]
	\e \frac1{4(d-2)!}\int d^{3}x\,\bigg[\,
	\frac1{(d-2)!}\left(\epsilon_{\rho_{1}\cdots\rho_{d-2}\lambda(\mu}\,
	\partial^{\lambda}\,\wh\varphi^{\,\rho_{1}\cdots \rho_{d-2},}{}_{\nu)}
	\right)^{2}\nn
	&&
       -\,\Big(\partial^{\nu}\,\wh\varphi_{\mu_{1}\cdots\mu_{d-2},\nu}\Big)
	\Big(\partial_{\rho}\,\wt\varphi^{\,\mu_{1}\cdots \mu_{d-2}\rho}\Big)
	-\frac12\,\Big(\partial_{\rho}\,\wt\varphi^{\,\mu_{1}\cdots \mu_{d-2}\rho}
	\Big)^{2}\,
	\bigg]\,,
\ea
and its $\wt\varphi$-shell condition:
\be
	\partial_{[\rho}\partial^{\nu}\,\wt\varphi_{\mu_{1}\cdots\mu_{d-2}]\nu}+
	\partial_{[\rho}\partial^{\nu}\,\wh\varphi_{\mu_{1}\cdots\mu_{d-2}],\nu}
	=0\,,
\ee
admits the following solution:
\ba
& \partial^{\nu}\,\wt\varphi_{\mu_{1}\cdots\mu_{d-2}\nu}
+ \partial^{\nu}\,\wh\varphi_{\mu_{1}\cdots\mu_{d-2},\nu}
=\partial_{[\mu_{1}}\,\chi_{\mu_{2}\cdots\mu_{d-2}]}\,,\\
&\partial^{\nu}\,\partial_{[\nu}\,\chi_{\mu_{1}\cdots\mu_{d-3}]}
=\partial^{\rho}\,\partial^{\sigma}\,
\wh\varphi_{\mu_{1}\cdots\mu_{d-3}\rho,\sigma}\,.
\ea
Analogously to the three dimensional case,
one can show that
the action \eqref{FPvb d} is (on-shell) equivalent to the action:
\ba\label{FPvad}
	S_{2}\big[\wh\varphi_{\mu_{1}\cdots\mu_{d-2},\nu},
	\chi_{\mu_{1}\cdots\mu_{d-3}}\big]\e
	\frac1{4(d-2)!}\int d^{d}x\,\bigg[\,
	\frac1{(d-2)!}\,\big(\epsilon_{\rho_{2}\cdots\rho_{d}(\mu}\,
	\partial^{\rho_{2}}\,\wh\varphi^{\,\rho_{3}\cdots \rho_{d}}{}_{\nu)}
	\big)^{2} \nn
	&&\hspace{50pt}
       +\,\frac{1}2\left(
       \partial_{[\mu_{1}}\,\chi_{\mu_2\cdots\mu_{d-2}]}
                -\partial^{\nu}\,\wh\varphi_{\mu_1\cdots\mu_{d-2},\nu}\right)^{2}\bigg]\,,
\ea
possessing the gauge symmetry given by \eqref{dual phi sym d} and
\ba
	\delta\,\chi_{\mu_{1}\cdots \mu_{d-3}}
	=\frac23\,(d-2) \left(\partial^{\nu}\,
	\wh\vartheta_{\mu_{1}\cdots\mu_{d-3},\nu}
	+\partial^{\nu}\,
	\wt\vartheta_{\mu_{1}\cdots\mu_{d-3}\nu}\right).
\ea
The following combination of $\wh\varphi_{\mu_{1}\cdots\mu_{d-2},\nu}$
and $\chi_{\mu_{1}\cdots\mu_{d-3}}$\,:
\be
	\varphi_{\mu_{1}\cdots\mu_{d-2},\nu}:=
	\wh\varphi_{\mu_{1}\cdots\mu_{d-2},\nu}
	+\frac12\,\eta_{\nu[\mu_{d-2}}\,
	\chi_{\mu_{1}\cdots\mu_{d-3}]}\,,
\ee
leads to the usual gauge transformation:
\be
	\delta\,\varphi_{\mu_{1}\cdots\mu_{d-2},\nu}
	=(d-2)\left[
	\partial_{[\mu_{d-2}}\,\wh\vartheta_{\mu_{1}\cdots\mu_{d-3}],\nu}
	+\partial_{[\mu_{d-2}}\,\wt\vartheta_{\mu_{1}\cdots\mu_{d-3}]\nu}
	-\partial_{[\mu_{d-2}}\,\wt\vartheta_{\mu_{1}\cdots\mu_{d-3}\nu]}
	\right].
\ee
Therefore, the two-derivative part of the action
becomes the (generalized) Einstein action for the hook field \cite{Curtright:1980yk} $\varphi_{\mu_{1}\cdots\mu_{d-2},\nu}$\,:
\be
	S_{\sst\rm M}[h_{\mu\nu}(\wh\varphi,\wt\varphi)]
	\approx S_{\sst\rm E}[\varphi_{\mu_{1}\cdots\mu_{d-2},\nu}]
	:=\frac1{4\,(d-2)!}\int d^{d}x\,
	\varphi^{\mu_{1}\cdots\mu_{d-2},\nu}\,G_{\mu_{1}\cdots\mu_{d-2},\nu}\,.
\ee
Finally, the total action reads
\be\label{NMG d}
	S[\varphi_{\mu_{1}\cdots\mu_{d-2},\nu}]=S_{\sst\rm B}[\varphi_{\mu_{1}\cdots\mu_{d-2},\nu}]
	-m^{2}\,S_{\sst\rm E}[\varphi_{\mu_{1}\cdots\mu_{d-2},\nu}]\,,
\ee
and it generalizes the NMG actions \cite{Bergshoeff:2009hq,Bergshoeff:2012ud} to arbitrary dimensions. In the massless limit, one ends up with the Bach action,
which propagates a massless spin two and a scalar (see Appendix \ref{sec: app}
for the proof).

\section{New Topologically Massive Theories}
\label{sec: NTMG}

In this section,
we turn to the so-called New topologically massive theories,
and show how the actions of those theories can be obtained from the ordinary formulation.
Our analysis goes alongside the works \cite{Bekaert:1998yp,Henneaux:2004jw} whose main concern is the EM duality.
The difference, or novelty, of our method lies in
introducing the Hamiltonian
with respect to a spatial direction $z$ rather than time $x^{0}$\,.
This allows dimensional reduction on the $z$ direction,
at the same time increasing the number of derivatives by solving
Hamiltonian constraints.
We consider two types of fields: $p$-form fields and two-columns fields of height $p$\,.
For \mt{p=1}\,, they provide the Maxwell-Chern-Simons(CS) action
and NTMG, respectively.\footnote{
It is known that in three dimensions, due to existence of the CS term, there exists a Lagrangian description for one propagating mode of massive spin one. It can be described either by the first-order action \cite{Townsend:1983xs}, or by the second-order (Maxwell-CS) one \cite{Deser:1981wh}.
For the fields with spin greater than one, there are more than two
different actions which describe the propagation of a single massive mode in three dimensions \cite{Deser:1981wh,Deser:1984kw,Aragone:1986hm,
Deser:1990ay,Tyutin:1997yn,Carlip:2008jk,Dalmazi:2008zh,
Andringa:2009yc,Dalmazi:2009pm,Dalmazi:2010bf,Chen:2011vp,Chen:2011yx,Arias:2012rs,Dengiz:2012jb}.}

For the sake of brevity, from now on,
we use notation $\mu[p]$ for totally antisymmetric indices $\mu_{1}\cdots\mu_{p}$\,.

\subsection{$p$-form field}
\label{sec: p-form}

For a better understanding of the derivation of NTMG,
we first consider that of the topologically massive $p$-form action.
We essentially follow the work \cite{Bekaert:1998yp}
where (anti-)self-dual $p$-form action has been derived.

We begin with the action of a $p$-form field $\mathcal A_{m{\sst [p]}}$ in \mt{d+1} dimensions:
\ba
	&& \mathcal S[\mathcal A_{m{\sst [p]}}]=
	-\frac{1}{2\,(p+1)!}\int d^{d}x\,dz\,
	\mathcal F^{m{\sst [p+1]}}\,\mathcal F_{m{\sst [p+1]}}\nn
	&&=
    -\int d^{d}x\,dz
    \left[\,\frac{1}{2\,(p+1)!}\,\mathcal F^{\mu{\sst [p+1]}}\,\mathcal F_{\mu{\sst [p+1]}}
    +\frac{1}{2\,p!}\,\Big(\,\partial_z\mathcal A_{\mu{\sst [p]}}-
    p\,\partial_\mu \mathcal A_{d\mu{\sst[p-1]}}\,\Big)^{2}\,\right],\quad
\ea
where $\mathcal F_{m{\sst [p+1]}}$ is the field strength defined by
\be
	\mathcal F_{m{\sst [p+1]}}:=(p+1)!\,\partial_{m}\mathcal A_{m{\sst [p]}}\,.
\ee
We introduce the canonically conjugate field with respect to the coordinate $z$ as
\be
\pi^{\mu{\sst [p]}}
:=\frac{\delta \cal S}{\delta (\partial_z \mathcal A_{\mu{\sst[p]}})}=
-\partial_z\mathcal A^{\mu{\sst [p]}}+p\,\partial^\mu \mathcal A^{d\mu{\sst [p-1]}}\,,
\ee
then the  action can be written as
\ba\label{p form Ham}
{\cal S}[\mathcal A_{m{\sst [p]}},\pi^{\mu{\sst [p]}}]\e\int d^{d}x\,dz\, \bigg[\,
\frac{1}{p!}\,\pi^{\mu{\sst [p]}}\,\partial_z\mathcal A_{\mu{\sst [p]}}+
\frac{1}{2\,p!}\,\pi_{\mu{\sst [p]}}\,\pi^{\mu{\sst[p]}}
-\frac{1}{2\,(p+1)!}\,\mathcal F^{\mu{\sst [p+1]}}\,\mathcal F_{\mu{\sst [p+1]}}\nn
&&\hspace{60pt}
+\,\frac{1}{(p-1)!}\,\mathcal A_{d\mu{\sst [p-1]}}\,\partial_{\nu}\pi^{\nu\mu{\sst [p-1]}}\,\bigg]\,.
\ea
Note that the sign of ``momentum'' squared term is unusual since
the role of ``time'' is played by the space-like coordinate $z$\,.
Due to the $p$-form gauge symmetry, the action
\eqref{p form Ham} involves a Lagrange multiplier
$\mathcal A_{d\mu{\sst [p-1]}}$\,. The corresponding constraint,
\be\label{constr pi}
	\partial_{\mu}\pi^{\mu{\sst [p]}}=0\,,
\ee
can be solved by a totally antisymmetric field $\mathcal B^{\mu\sst [p+1]}$ as
\be\label{sol constr pi}
	\pi^{\mu{\sst [p]}}=\partial_{\nu}\,\mathcal B^{\mu{\sst [p]}\nu}\,.
\ee
At this point, we focus on the dimension \mt{d=2p+1}\,,
and dualize the solution field as
\be
	\mathcal B^{\mu\sst [p+1]}=
	\epsilon^{\mu{\sst [p+1]}\nu{\sst [p]}}\,\mathcal B_{\nu\sst [p]}\,.
\ee
In terms of the dual field $\mathcal B_{\nu\sst [p]}$, the action \eqref{p form Ham}
is given by
\ba \label{dual p form act}
{\cal S}[\mathcal A_{\mu{\sst [p]}},\mathcal B_{\mu{\sst [p]}}]\e
\int d^{2p+1}x\,dz\,\bigg[\,
\frac1{(p!)^{2}}\,\epsilon^{\mu{\sst [2p+1]}}\,
\partial_z\mathcal A_{\mu{\sst [p]}}
\,\partial_{\mu}\mathcal B_{\mu{\sst [p]}}\nn
&&-\,\frac1{2\,(p+1)!}\,
\mathcal J^{\mu{\sst [p+1]}}\,\mathcal J_{\mu{\sst [p+1]}}-\frac1{2\,(p+1)!}\,
\mathcal F^{\mu{\sst [p+1]}}\,\mathcal F_{\mu{\sst [p+1]}}\bigg]\,,
\ea
where \mt{\mathcal J_{\mu\sst [p+1]}:=
(p+1)!\,\partial_{\mu}\mathcal B_{\mu\sst [p]}} is the field strength of $B_{\mu\sst [p]}$\,.

\subsubsection*{(Anti-)Self-dual $p$-form (\mt{p=2k})}

Let us focus on the first term of the action \eqref{dual p form act}:
\be
	\mathcal P[\mathcal A_{\mu{\sst [p]}},\mathcal B_{\mu{\sst [p]}}]
	:=\frac1{(p!)^{2}}\int d^{2p+1}x\,dz\,
	\epsilon^{\mu{\sst [2p+1]}}\,
	\partial_z\mathcal A_{\mu{\sst [p]}}\,
	\partial_{\mu}\mathcal B_{\mu{\sst [p]}}\,,
\ee
which has the following symmetry property:
\be\label{P d+1}
	\mathcal P[\mathcal A_{\mu{\sst [p]}},\mathcal B_{\mu{\sst [p]}}]=
	(-1)^{p}\,\mathcal P[\mathcal B_{\mu{\sst [p]}},\mathcal A_{\mu{\sst [p]}}]\,.
\ee
For even $p=2k$ (with dimensions \mt{d+1=4k+2}),
it becomes symmetric so that
one can decompose the action into those of self-dual and anti-self-dual $p$-form:
\be
	\mathcal S[\mathcal A_{\mu{\sst [p]}},\mathcal B_{\mu{\sst [p]}}]
	=\mathcal S_{\sst\rm +C}[\mathcal A^+_{\mu{\sst [p]}}]
	+\mathcal S_{\sst\rm -C}[\mathcal A^-_{\mu{\sst [p]}}]\,,
	\qquad
	\mathcal A^{\pm}_{\mu{\sst [p]}}=\frac1{\sqrt2}\left(
	\mathcal A_{\mu{\sst [p]}}\pm \mathcal B_{\mu{\sst [p]}}\right).
\ee
Here, the action of  (anti-)self-dual $p$-form \cite{Henneaux:1988gg} reads
\be
	{\cal S}_{\sst\rm \pm C}[\mathcal A_{\mu{\sst [p]}}]=\frac12
	\int d^{2p+1}x\,dz
	\left[\,\pm\frac1{(p!)^{2}}\,\epsilon^{\mu{\sst [2p+1]}}\,
	\partial_z\mathcal A_{\mu{\sst [p]}}\,
	\partial_{\mu}\mathcal A_{\mu{\sst [p]}}
	-\frac1{(p+1)!}\,
	\mathcal F^{\mu{\sst [p+1]}}\,\mathcal F_{\mu{\sst [p+1]}}
	\right].
\ee

\subsubsection*{Topologically massive $p$-form (\mt{p=2k+1})}

When the form-degree is odd: \mt{p=2k+1},
that is when \mt{d=4k+3}\,,
one can consider the dimensional reduction:
\be\label{dimredp}
	\mathcal A_{\mu{\sst [p]}}(x,z)= \cos(m\,z)\,A_{\mu{\sst [p]}}(x)\,,
	\qquad
	\mathcal B_{\mu{\sst [p]}}(x,z)= \sin(m\,z)\,B_{\mu{\sst [p]}}(x)\,.
\ee
Then, the first term of the action \eqref{dual p form act}
gives
\be
	P[A_{\mu{\sst [p]}},B_{\mu{\sst [p]}}]:=\frac1{(p!)^{2}}\int d^{2p+1}x\,
	\epsilon^{\mu{\sst [2p+1]}}\,
	A_{\mu{\sst [p]}}\,\partial_{\mu}B_{\mu{\sst [p]}}\,,
\ee
which has a different symmetry property compared to \eqref{P d+1}:
\be
	P[A_{\mu{\sst [p]}},B_{\mu{\sst [p]}}]=(-1)^{p+1}\,P[B_{\mu{\sst [p]}},A_{\mu{\sst [p]}}]\,.
\ee
Hence, it becomes symmetric for \mt{p=2k+1}, and
the dimensionally reduced action  can be again split into two copies of
topologically massive action with opposite mass:
\be
	S[A_{\mu{\sst [p]}}\,,B_{\mu{\sst [p]}}]=
	S_{\sst\rm T}[+m; A^{+}_{\mu{\sst [p]}}]+
	S_{\sst\rm T}[-m; A^{-}_{\mu{\sst [p]}}]\,,
\qquad
	A^{\pm}_{\mu{\sst [p]}}=\frac1{\sqrt2}\left(
	A_{\mu{\sst [p]}}\pm B_{\mu{\sst [p]}}\right),
\ee
where $S_{\sst\rm T}$ is the $p$-form
generalization of the Maxwell-CS action:
\be
	S_{\sst\rm T}[m;A_{\mu{\sst [p]}}]= -\frac12\int d^{2p+1}x
	\left[\,\frac1{(p+1)!}\,
	F^{\mu{\sst [p+1]}}\,F_{\mu{\sst [p+1]}}
	+\frac{m}{(p!)^{2}}\,\epsilon^{\mu{\sst [2p+1]}}\,
	A_{\mu{\sst [p]}}\,
	\partial_{\mu}A_{\mu{\sst [p]}}\right].
\ee

\subsection{Field with two-column Young symmetry}

Let us consider now the fields with two-column Young symmetry:
\be
	\mathcal H \ :\ {\scriptsize{\left.\yng(2,2,2)\right\}p}}\ ,
\ee
or, the $[p,p]$-symmetry fields.\footnote{
In this paper, the $[p,q]$-symmetry refers to  the index symmetry of
 two-column Young diagram with respective height $p$ and $q$\,.}
We begin with the \mt{(d+1)}-dimensional massless action
\cite{Curtright:1980yk}
for $[p,p]$-symmetry field:
\be\label{pp action}
	\mathcal S[\mathcal H^{n{\sst [p]}}_{m{\sst [p]}}]=
	-\frac1{4\,(p!)^{2}}\int d^{d}x\,dz\,\delta^{m{\sst [2p+1]}}_{n{\sst [2p+1]}}\,
	\partial_{m}\,\mathcal H^{n{\sst [p]}}_{m{\sst [p]}}\,
	\partial^{n}\,\mathcal H^{n{\sst [p]}}_{m{\sst [p]}}\,,
\ee
where $\delta^{m\sst [p]}_{n \sst[p]}$ is the generalized Kronecker delta:
\be
	\delta^{m\sst [p]}_{n \sst[p]}=
	\delta^{m_{1}\cdots m_{p}}_{n_{1}\cdots n_{p}}:=
	p!\,
	\delta^{[m_{1}}_{[n_{1}}\cdots \delta^{m_{p}]}_{n_{p}]}\,.
\ee
This action describes a massless particle carrying  the helicity representation of $[p,p]$ Young diagram, and it is invariant under the gauge transformations:
\be
	\delta \mathcal H^{m\sst [p]}_{n\sst [p]}
	=\partial_{n}\,\xi^{m\sst [p]}_{n\sst [p-1]}+\partial^{m}\xi^{m\sst [p-1]}_{n\sst [p]}\,,
\ee
with the $[p,p-1]$-symmetry parameter $\xi^{m\sst [p]}_{n \sst [p-1]}$\,.

For our purpose, we first write the action \eqref{pp action}
in a way that the spatial direction $z$ is distinguished:
\ba\label{pp action2}
	 \mathcal S[\mathcal H^{n{\sst [p]}}_{m{\sst [p]}}]\e
	\int d^{d}x\,dz\,\bigg[\,
	\frac{-1}{4\,(p!)^{2}}\,\delta^{\mu{\sst [2p+1]}}_{\nu{\sst [2p+1]}}\,
	\partial_{\mu}\,\mathcal H^{\nu{\sst [p]}}_{\mu{\sst [p]}}\,
	\partial^{\nu}\,\mathcal H^{\nu{\sst [p]}}_{\mu{\sst [p]}}
	+\frac{(-1)^{p}}{2\,[(p-1)!]^{2}}\,\delta^{\mu{\sst [2p]}}_{\nu{\sst [2p]}}\,
	\mathcal H^{d\nu{\sst [p-1]}}_{d\mu{\sst [p-1]}}\,
	\partial_{\mu}\,\partial^{\nu}\,\mathcal H^{\nu{\sst [p]}}_{\mu{\sst [p]}}
	\nn
	&&\hspace{50pt}
	+\,\frac{(-1)^{p+1}}{4\,(p!)^{2}}\,\delta^{\mu{\sst [2p]}}_{\nu{\sst [2p]}}
	\left(\partial_{z}\mathcal H^{\nu{\sst [p]}}_{\mu{\sst [p]}}
	-p\,\partial_{\mu}\mathcal H_{d\mu{\sst [p-1]}}^{\nu{\sst [p]}}
	-p\,\partial^{\nu}\mathcal H_{\mu{\sst [p]}}^{d\nu{\sst [p-1]}}\right)^{2}\bigg]\,.
\ea
Then, we introduce the canonically conjugate field as
\be
	 \pi^{\rho{\sst [p]}}_{\sigma{\sst [p]}}:=
	 \frac{\delta \cal S}{\delta (\partial_z \mathcal H^{\sigma{\sst [p]}}_{\rho{\sst[p]}})}
	 =\delta^{\mu{\sst [p]}\rho{\sst [p]}}_{\sigma{\sst [p]}\nu{\sst [p]}}
	 \left(-\partial_{z}\mathcal H_{\mu{\sst [p]}}^{\nu{\sst [p]}}
	 +p\,\partial^{\nu}\mathcal H_{\mu{\sst [p]}}^{d\nu{\sst [p-1]}}
	 +p\,\partial_{\mu}\mathcal H_{d\mu{\sst [p-1]}}^{\nu{\sst [p]}}\right).
\ee
and recast the action \eqref{pp action2} into the Hamiltonian one as
\ba\label{pp Ham act}
	\mathcal S[\mathcal H^{n{\sst [p]}}_{m{\sst [p]}},\pi^{\mu{\sst [p]}}_{\nu{\sst [p]}}]\e
	\int d^{d}x\,dz\,\bigg[\,
	\frac1{2\,(p!)^{2}}\,
	\pi^{\mu{\sst [p]}}_{\nu{\sst [p]}}\,\partial_{z}\mathcal H^{\nu{\sst [p]}}_{\mu{\sst [p]}}\nn
	&&
	+\frac1{4\,(p!)^{2}}\,\gamma^{\nu{\sst [2p]}}_{\mu{\sst [2p]}}\,\pi^{\mu{\sst [p]}}_{\nu{\sst [p]}}\,\pi^{\mu{\sst [p]}}_{\nu{\sst [p]}}\,
	-\frac1{4\,(p!)^{2}}\,\delta^{\mu{\sst [2p+1]}}_{\nu{\sst [2p+1]}}\,
	\partial_{\mu}\mathcal H^{\nu{\sst [p]}}_{\mu{\sst [p]}}\,\partial^{\nu}
	\mathcal H^{\nu{\sst [p]}}_{\mu{\sst [p]}}	\\
	&&
	-\,\frac1{2\,(p-1)!\,p!}\,\mathcal H^{\nu{\sst [p-1]}d}_{\mu{\sst[p]}}
	\,\partial^{\nu}\pi^{\mu{\sst [p]}}_{\nu{\sst [p]}}
	+\frac{(-1)^{p}}{2\,[(p-1)!]^{2}}\,\delta^{\mu{\sst [2p]}}_{\nu{\sst [2p]}}\,
	\mathcal H^{\nu{\sst [p-1]}d}_{\mu{\sst[p-1]}d}\,\partial_{\mu}\partial^{\nu}\mathcal H^{\nu{\sst [p]}}_{\mu{\sst [p]}}\,
	\bigg]\,.\nonumber
\ea
Here, $\gamma^{\nu{\sst [2p]}}_{\mu{\sst [2p]}}$ is the inverse   of
$\delta^{\mu{\sst [2p]}}_{\nu{\sst [2p]}}$ defined by
\be
	\delta^{\mu{\sst[p]}\rho{\sst[p]}}_{\sigma{\sst[p]}\nu{\sst[p]}}\,
	\gamma^{\sigma{\sst [p]}\kappa{\sst[p]}}_{\rho{\sst[p]}\lambda{\sst[p]}}
	=\frac1{(p!)^{2}}\,\delta^{\mu{\sst[p]}}_{\lambda{\sst[p]}}\,
	\delta^{\kappa{\sst[p]}}_{\nu{\sst[p]}}\,.
\ee
The last line of the action \eqref{pp Ham act} involves
two Lagrangian multipliers and the system is subject to the corresponding constraints:
\be\label{ConstR}
	\partial^{\nu}\pi^{\mu{\sst [p]}}_{\nu{\sst [p]}}=0\,,
	\qquad
	\delta^{\mu{\sst [2p]}}_{\nu{\sst [2p]}}\,
	\partial_{\mu}\partial^{\nu}\mathcal H^{\nu{\sst [p]}}_{\mu{\sst [p]}}=0\,.
\ee
By solving these constraints, one can transform the action  \eqref{pp Ham act}
into a higher-derivative one.
\begin{itemize}
\item
First, the constraint on the conjugate field can be solved as
\be\label{SC1}
	\pi^{\mu{\sst [p]}}_{\nu{\sst [p]}}
	=\partial_{\rho}\partial^{\sigma}
	\mathcal U^{\rho\mu{\sst [p]}}_{\sigma\nu{\sst[p]}}\,,
\ee
where $\mathcal U^{\mu\sst [p+1]}_{\nu\sst [p+1]}$ is a \mt{[p+1,p+1]}-symmetry field.
When \mt{d=2p+1}, the traceless part of $\mathcal U^{\mu\sst [p+1]}_{\nu\sst [p+1]}$
vanishes identically and it is equivalent to its trace part, that is a $[p,p]$-symmetry field:
\be
	\mathcal U^{\mu{\sst [p+1]}}_{\nu{\sst[p+1]}}
	=\delta^{\mu{\sst[p+1]}\rho{\sst [p]}}_{\nu{\sst [p+1]}\sigma{\sst[p]}}\,
	\mathcal U^{\sigma{\sst[p]}}_{\rho{\sst[p]}}\,.
\ee
In terms of  $\mathcal U^{\sigma{\sst[p]}}_{\rho{\sst[p]}}$
(which can be also viewed as the double dual of $\mathcal U^{\mu{\sst [p+1]}}_{\nu{\sst[p+1]}}$),
the solution for the conjugate field has the form of the (generalized) Einstein tensor
$G^{\mu{\sst[p]}}_{\nu{\sst[p]}}$ for $[p,p]$-symmetry field:
\be\label{Sol1}
	\pi^{\mu{\sst[p]}}_{\nu{\sst[p]}}=(-1)^{p}\,
	G^{\mu{\sst[p]}}_{\nu{\sst[p]}}(\mathcal U)=
	\delta^{\mu{\sst[p]}\rho{\sst [p+1]}}_{\nu{\sst [p]}\sigma{\sst[p+1]}}\,
	\partial_{\rho}\partial^{\sigma}\mathcal U^{\sigma{\sst[p]}}_{\rho{\sst[p]}}\,.
\ee
\item
The second constraint in \eqref{ConstR} can be solved as
\be
	\mathcal H^{\nu{\sst [p]}}_{\mu{\sst [p]}}=\frac12\,
	\gamma^{\nu{\sst[p]}\sigma\sst[p]}_{\mu{\sst[p]}\rho\sst [p]}
	\left(\partial_{\lambda}\mathcal V^{\lambda\rho{\sst[p]}}_{\sigma{\sst[p]}}
	+\partial^{\lambda}\mathcal V^{\rho{\sst[p]}}_{\lambda\sigma{\sst[p]}}\right),
\ee
where $\mathcal V^{\nu\sst [p+1]}_{\mu\sst [p]}$ is a $[p+1,p]$-symmetry field.
After dualizing  $\mathcal V^{\nu\sst [p+1]}_{\mu\sst [p]}$\,, the solution can be recast into
\be\label{Sol2}
	\mathcal H^{\nu{\sst [p]}}_{\mu{\sst [p]}}=
	\frac12\left(\epsilon_{\rho{\sst [p+1]}\mu{\sst [p]}}\,\partial^{\rho}
	\mathcal V^{\rho{\sst [p]},\nu{\sst [p]}}+
	\epsilon^{\sigma{\sst [p+1]}\nu{\sst [p]}}\,\partial_{\sigma}\,
	\mathcal V_{\mu{\sst [p]},\sigma{\sst [p]}}
	+ \partial_{\mu}\,\mathcal W^{\nu{\sst [p]}}_{\mu{\sst [p-1]}}
	+\partial^{\nu}\, \mathcal W^{\nu{\sst [p-1]}}_{\mu{\sst [p]}}\right),
\ee
where $\mathcal W^{\nu{\sst [p]}}_{\mu{\sst [p-1]}}$
is a $[p,p-1]$-symmetry field given by
the trace part of $\mathcal V^{\nu{\sst [p+1]}}_{\mu{\sst [p]}}$.
$\mathcal V_{\mu{\sst [p]},\nu{\sst [p]}}$
is the dual of the traceless part of $\mathcal V^{\nu{\sst [p+1]}}_{\mu{\sst [p]}}$
so it has $[p,p]$-symmetry.
Since the trace part of $\mathcal V_{\mu{\sst [p]},\nu{\sst [p]}}$
does not contribute to the expression \eqref{Sol2}, we consider henceforth
$\mathcal V_{\mu{\sst [p]},\nu{\sst [p]}}$ as traceful.
\end{itemize}
We substitute the solutions \eqref{Sol1} and \eqref{Sol2} in
the action \eqref{pp Ham act}.
Notice first that, since $\pi^{\mu{\sst[p]}}_{\nu{\sst[p]}}$ is given by the Einstein tensor,
the action is invariant under the gauge transformation of
$\mathcal H^{\nu{\sst [p]}}_{\mu{\sst [p]}}$\,.
Consequently, the $\mathcal W^{\nu{\sst [p]}}_{\mu{\sst [p-1]}}$ terms in the solution \eqref{Sol2}
do not contribute, and the action becomes a functional of two
$[p,p]$-symmetry fields
$\mathcal U^{\nu{\sst [p]}}_{\mu{\sst [p]}}$ and $\mathcal V^{\nu{\sst [p]}}_{\mu{\sst [p]}}$\,.
All in all, the resulting action reads
\ba\label{pp NT}
	\mathcal S[\mathcal U^{\nu{\sst [p]}}_{\mu{\sst [p]}},
	\mathcal V^{\nu{\sst [p]}}_{\mu{\sst [p]}}]
	\e
	\frac1{(p!)^{2}}\int d^{2p+1}x\,dz\,\bigg[\,-\frac12
	\epsilon^{\mu\sst [2p+1]}\,\partial_{z}
	\mathcal V_{\mu{\sst [p]},\nu{\sst [p]}}\,\partial_{\mu}
	G^{\nu{\sst [p]}}_{\mu{\sst [p]}}(\mathcal U) \nn
	&&\hspace{60pt}
	+\,\frac14\,G^{\nu{\sst [p]}}_{\mu{\sst [p]}}(\mathcal U)\,
	S^{\mu{\sst [p]}}_{\nu{\sst [p]}}(\mathcal U)
	+\frac14\,G^{\nu{\sst [p]}}_{\mu{\sst [p]}}(\mathcal V)\,
	S^{\mu{\sst [p]}}_{\nu{\sst [p]}}(\mathcal V)\,\bigg]\,,
\ea
where $S^{\mu{\sst [p]}}_{\nu{\sst [p]}}$ is the $[p,p]$-symmetry
generalization of the Schouten tensor:
\be
	S^{\nu{\sst[p]}}_{\mu{\sst[p]}}
	:=\gamma^{\nu{\sst [p]}\sigma{\sst [p]}}_{\mu{\sst [p]}\rho{\sst [p]}}\,
	G^{\rho{\sst[p]}}_{\sigma{\sst[p]}}\,.
\ee
Let us make a few comments on the expression \eqref{pp NT}.
The first and second terms are straightforward result of the substitution.
They are invariant under the gauge plus Weyl transformation:
\be\label{pp gauge tr}
	\delta\,\mathcal U^{\mu{\sst[p]}}_{\nu{\sst[p]}}
	=\partial^{\mu}\xi^{\mu{\sst[p-1]}}_{\nu{\sst[p]}}
	+\partial_{\nu}\xi^{\mu{\sst[p]}}_{\nu{\sst[p-1]}}
	+\delta^{\mu}_{\nu}\,\alpha^{\mu{\sst[p-1]}}_{\nu{\sst[p-1]}}\,,
\ee
due to the property of the Schouten tensor:
\be
	S^{\nu{\sst[p]}}_{\mu{\sst[p]}}(\delta\,\mathcal U)=
	\partial_{\mu}\partial^{\nu}\alpha^{\nu{\sst[p-1]}}_{\mu{\sst[p-1]}}\,.
\ee
In fact, they are unique functionals, up to factors, invariant under
the transformation \eqref{pp gauge tr}
and involving three and four derivatives, respectively.
One can see as well that the third term of \eqref{pp Ham act} gives that of \eqref{pp NT},
by examining their symmetries.
In eq.~\eqref{Sol2},
the solution for $\mathcal H^{\nu{\sst [p]}}_{\mu{\sst [p]}}$ is invariant under
\ba\label{V trf}
	&&\delta\,\mathcal V^{\mu{\sst[p]}}_{\nu{\sst[p]}}
	=\partial^{\mu}\xi^{\mu{\sst[p-1]}}_{\nu{\sst[p]}}
	+\partial_{\nu}\xi^{\mu{\sst[p]}}_{\nu{\sst[p-1]}}
	+\delta^{\mu}_{\nu}\,\alpha^{\mu{\sst[p-1]}}_{\nu{\sst[p-1]}}\,,\\
	&&\delta\,\mathcal W^{\mu\sst [p]}_{\nu\sst [p-1]}
	=-\epsilon_{\rho\sst [p+1]}{}^{\mu\sst[p]}\,
	\partial^{\rho}\,\xi^{\rho{\sst[p]}}_{\nu{\sst[p-1]}}\,.
\ea
Since $\mathcal \mathcal \mathcal W^{\mu\sst [p]}_{\nu\sst [p-1]}$
decouples from the action, the third term of \eqref{pp Ham act}
is invariant under the transformation \eqref{V trf}\,,
so is necessarily proportional to the last term of \eqref{pp NT},
that is the Bach action.
The overall constant can be easily fixed by comparing the
\mt{\mathcal V^{\nu{\sst[p]}}_{\mu{\sst[p]}}\,\Box^{2}\,
\mathcal V^{\mu{\sst[p]}}_{\nu{\sst[p]}}} terms.

\subsubsection*{(Anti-)Self-dual $[p,p]$-symmetry field (\mt{p=2k})}

Let us consider the first term of the action \eqref{pp NT}:
\be
	\mathcal P
	[\mathcal U^{\nu{\sst [p]}}_{\mu{\sst [p]}},\mathcal V^{\nu{\sst [p]}}_{\mu{\sst [p]}}]:=
	\frac1{(p!)^{2}}\int d^{2p+1}x\,dz\
	\epsilon^{\mu\sst [2p+1]}\,\partial_{z}\mathcal V_{\mu{\sst [p]},\nu{\sst [p]}}\,
	\partial_{\mu}G^{\nu{\sst [p]}}_{\mu{\sst [p]}}(\mathcal U)\,,
\ee
which has the symmetry property:
\be
	\mathcal P[\mathcal U^{\nu{\sst [p]}}_{\mu{\sst [p]}},
	\mathcal V^{\nu{\sst [p]}}_{\mu{\sst [p]}}]
	=(-1)^{p}\,\mathcal P[\mathcal V^{\nu{\sst [p]}}_{\mu{\sst [p]}},
	\mathcal U^{\nu{\sst [p]}}_{\mu{\sst [p]}}]\,.
\ee
For even \mt{p=2k} (with dimensions \mt{d+1=4k+2}), it becomes symmetric
and
one can decompose the action into those of self-dual and anti-self-dual fields:
\be
	\mathcal S[\mathcal U^{\nu{\sst [p]}}_{\mu{\sst [p]}},\mathcal
	V^{\nu{\sst [p]}}_{\mu{\sst [p]}}]
	=\mathcal S_{\sst\rm +C}[\Phi^{+\,\nu{\sst [p]}}_{\mu{\sst [p]}}]
	+\mathcal S_{\sst\rm -C}[\Phi^{-\,\nu{\sst [p]}}_{\mu{\sst [p]}}]\,,
	\qquad
	\Phi^{\pm\,\nu{\sst [p]}}_{\mu{\sst [p]}}=\frac1{\sqrt2}\left(
	\mathcal V^{\nu{\sst [p]}}_{\mu{\sst [p]}}\pm
	\mathcal U^{\nu{\sst [p]}}_{\mu{\sst [p]}}\right).
\ee
The action of  (anti-)self-dual $[p,p]$-symmetry field reads
\be
	\mathcal S_{\sst\rm \pm C}[\Phi^{\nu{\sst [p]}}_{\mu{\sst [p]}}]
	=
	\frac1{2(p!)^{2}}\int d^{2p+1}x\,dz\,\bigg[\,
	\mp\,\epsilon^{\mu\sst [2p+1]}\,\partial_{z}
	\Phi_{\mu{\sst [p]},\nu{\sst [p]}}\,
	\partial_{\mu}G^{\nu{\sst [p]}}_{\mu{\sst [p]}}(\Phi)
	+G^{\nu{\sst [p]}}_{\mu{\sst [p]}}(\Phi)\,
	S^{\mu{\sst [p]}}_{\nu{\sst [p]}}(\Phi)\,\bigg]\,.
\ee	

\subsubsection*{Topologically massive $[p,p]$-symmetry field (\mt{p=2k+1})}

If we perform the dimensional reduction:
\be
	\mathcal U^{\nu\sst [p]}_{\mu{\sst [p]}}(x,z)=
	\cos(m\,z)\,U^{\nu\sst [p]}_{\mu{\sst [p]}}(x)\,,
	\qquad
	\mathcal V_{\mu{\sst [p]}}(x,z)= \sin(m\,z)\,V^{\nu\sst [p]}_{\mu{\sst [p]}}(x)\,.
\ee
then, the first term of the action \eqref{pp NT} gives
\be
	P[U^{\nu{\sst [p]}}_{\mu{\sst [p]}},V^{\nu{\sst [p]}}_{\mu{\sst [p]}}]:=
	\frac1{(p!)^{2}}\int d^{2p+1}x\,
	\epsilon^{\mu\sst [2p+1]}\,V_{\mu{\sst [p]},\nu{\sst [p]}}\,
	\partial_{\mu}G^{\nu{\sst [p]}}_{\mu{\sst [p]}}(U)\,.
\ee
satisfying
\be
	P[U^{\nu\sst [p]}_{\mu{\sst [p]}},V^{\nu\sst [p]}_{\mu{\sst [p]}}]
	=(-1)^{p+1}\,P[V^{\nu\sst [p]}_{\mu{\sst [p]}},U^{\nu\sst [p]}_{\mu{\sst [p]}}]\,.
\ee
When \mt{p=2k+1} (with \mt{d=4k+3}), the above functional
becomes symmetric. Consequently,
the dimensionally reduced massive action
can be split into two copies of topologically massive action with opposite mass:
\be
	S[U^{\nu\sst [p]}_{\mu{\sst [p]}}\,,V^{\nu\sst [p]}_{\mu{\sst [p]}}]=
	S_{\sst\rm NTMG}[+m; \phi^{+\,\nu\sst [p]}_{\mu{\sst [p]}}]+
	S_{\sst\rm NTMG}[-m; \phi^{-\,\nu\sst [p]}_{\mu{\sst [p]}}]\,,
\qquad
	\phi^{\pm\,\nu\sst [p]}_{\mu{\sst [p]}}=\frac1{\sqrt2}\left(
	V^{\nu\sst [p]}_{\mu{\sst [p]}}\pm U^{\nu\sst [p]}_{\mu{\sst [p]}}\right),
\ee
with
\be
	S_{\sst\rm NTMG}[m;\phi^{\nu{\sst [p]}}_{\mu{\sst [p]}}]
	=
	\frac1{2(p!)^{2}}\int d^{d}x\,\bigg[\,
	m\,\epsilon^{\mu\sst [2p+1]}\,\phi_{\mu{\sst [p]},\nu{\sst [p]}}\,
	\partial_{\mu}\,G^{\nu{\sst [p]}}_{\mu{\sst [p]}}(\phi)
	+G^{\nu{\sst [p]}}_{\mu{\sst [p]}}(\phi)\,
	S^{\mu{\sst [p]}}_{\nu{\sst [p]}}(\phi)\,\bigg]\,.
\ee	
This action $S_{\sst\rm NTMG}$ generalizes NTMG (which corresponds to \mt{p=1} case)
to $[p,p]$-symmetry field,
and describes half of the helicity states carrying the $[p,p]$-symmetry representation.

\section{Conclusions}
\label{sec: disc}

In the present paper, we have shown how a class of higher-derivative massive theories
with unitary propagation can be obtained from dimensional reduction of
ordinary massless actions.
The procedure used here for NMG and NTMG can be
also applied to higher spins, and
it would be interesting to compare the results in three dimensions
with the actions recently obtained in \cite{Bergshoeff:2009tb,Bergshoeff:2011pm,Chen:2011vp,Chen:2011yx}.
It is also tempting to speculate that,
in the case of three dimensional higher spins,
there may exist more than one higher-derivative massive theories
which make use of the hierarchy of actions derived in \cite{Joung:2012qy}
(see also \cite{Francia:2012rg}).
Hopefully, we will report about these issues in the near future.

\acknowledgments

We would like to thank Dario Francia for discussions on the subject of this work and Augusto Sagnotti for constant encouragement.
This work was supported in part by Scuola Normale Superiore, by INFN (I.S. TV12) and by the MIUR-PRIN contract 2009-KHZKRX. The work of KM is also supported by the ERC Advanced Investigator Grants no. 226455 ÒSupersymmetry, Quantum Gravity and Gauge FieldsÓ (SUPERFIELDS).
	
\appendix

\section{Appendix}
\label{sec: app}

In this Appendix, we consider the massless limit of the actions derived in this paper, and show their connection to other unconventional actions known in the literature.

In the case of spin one, starting from the action \eqref{p form Ham} with \mt{p=1}, one can solve the constraint \eqref{constr pi} as \eqref{sol constr pi}
introducing an antisymmetric field $B^{\mu\nu}$.
After dimensional reduction (but without dualization), one ends up with  the action:
\be\label{spin 1 B}
S[A_{\mu},B^{\mu\nu}]=\int d^{d}x\, \left[\frac{1}{2}\,\partial_{\lambda}B^{\mu\lambda}\partial^{\nu}B_{\mu\nu}
-\frac{1}{4}F_{\mu\nu}F^{\mu\nu}+m\,\partial^{\nu}B_{\mu\nu}A^{\mu}\right]\,.
\ee
In the massless limit, this action does not contain any mixing term.
The $A_{\mu}$ part is given by a Maxwell action, therefore describes a massless spin one,
whereas the $B^{\mu\nu}$ part corresponds to the so-called
``notoph'' action \cite{Ogievetsky:1967ij}, describing a massless scalar.

Analogously, in the case of spin two,
starting from the action \eqref{pp Ham act} with \mt{p=1}, one can
perform dimensional reduction to get
\be\label{Ham 2}
	S[h_{mn},\pi^{\mu\nu}]= S_{1}[\pi^{\mu\nu},h_{\mu d}]+
	S_{2}[h_{\mu\nu},h_{dd}] +\frac{m}2\int d^{d}x\ \pi^{\mu\nu}\,h_{\mu\nu}\,,
\ee
where $S_{1}$ and $S_{2}$ are given by:
\ba
	&& S_{1}=S_{\sst\rm N}[\pi^{\mu\nu}]
	 -\frac12\int d^{d}x\,h_{\mu d}\,\partial_{\nu}\,\pi^{\mu\nu}\,,
	 \qquad
	 S_{\sst\rm N}[\pi^{\mu\nu}]:=\frac14\int d^{d}x\,\left(
	\pi^{\mu\nu}\,\pi_{\mu\nu}-\tfrac1{d-1}\,{\pi^{\mu}_{\ \mu}}^{2}
	\right),
	 \nn
	&& S_{2}=S_{\sst\rm E}[h_{\mu\nu}]
	 +\frac12\int d^{d}x\,h_{dd}\,(\partial^{\mu}\,\partial^{\nu}\,h_{\mu\nu}
	-\Box\,h^{\mu}_{\ \mu})\,.\label{S2}
\ea
In the massless $m=0$ limit, the action \eqref{Ham 2}
becomes a sum of two independent actions $S_{1}$ and $S_{2}$\,.

\paragraph{Spin one mode}

The action $S_{1}$ can be shown to be on-shell equivalent to the Maxwell action:
\be
	S_{1}=-\int d^{d}x\,\partial^{[\mu}\,A^{\nu]}\,
	\partial_{[\mu}\,A_{\nu]}\,,\qquad A_{\mu}:=h_{\mu d}\,,
\ee
after solving the equations of motion of the conjugate field $\pi_{\mu\nu}$.

On the other hand, one can solve the constraint as \eqref{SC1}
by introducing a field $U_{\mu\nu,\rho\sigma}$ of $[2,2]$-symmetry,
and get an equivalent four-derivative action:
\ba
	S_{1} \e S_{\sst\rm N}[\pi^{\mu\nu}(U)] \nn
	\e \frac14\int d^{d}x\,\left[
	U_{\mu\nu,\rho\sigma}\,\partial^{\rho}\,\partial^{\sigma}\,
	\partial_{\lambda}\,\partial_{\kappa}\,U^{\mu\nu,\lambda\kappa}
	-\tfrac1{d-1}\,
	U^{\mu}_{\ \ \mu,\rho\sigma}\,\partial^{\rho}\,\partial^{\sigma}\,
	\partial_{\lambda}\,\partial_{\kappa}\,U_{\nu}^{\ \ \nu,\lambda\kappa}
	\right].
\ea
This action propagates a massless spin one, so-called ``notivarg'' \cite{Deser:1980fy}.

\paragraph{Spin two and scalar modes}

The action $S_{2}$\, can be diagonalized as
\be\label{S2t}
	S_{2}=S_{\sst\rm E}[\tilde h_{\mu\nu}]
	+\frac12\int d^{d}x\,\phi\,\Box\,\phi\,,
	\qquad (\,\tilde h_{\mu\nu}\,,\,\phi\,):=
	(\,h_{\mu\nu}+\tfrac{1}{d-2}\eta_{\mu\nu}\,h_{dd}\,,\sqrt{2\,\tfrac{d-1}{d-2}}\,h_{dd}\,)\,,
\ee
making  obvious its propagating content.

Alternatively, one can also solve the constraint \eqref{Const} by \eqref{h(V)} and follow the dualization procedure of Section \ref{sec: NMG} to arrive at the four-derivative action:
\be
S_2 = S_{\sst\rm E}[h(\varphi)] = S_{\sst\rm B}[\varphi_{\mu_{1}\cdots\mu_{d-2},\nu}]\,.
\ee
This is the Bach action for the hook field $\varphi_{\mu_1\cdots\mu_{d-2},\nu}$\,,
with spin two and zero unitary propagating DoF, as one can see from the equivalent action \eqref{S2t}.

\bibliographystyle{JHEP}
\bibliography{hs_geometry.bib}

\end{document}